\documentclass[11pt]{article}

\usepackage[preprint]{acl}

\usepackage{times}
\usepackage{latexsym}

\usepackage[T1]{fontenc}
\usepackage[utf8]{inputenc}

\usepackage{inconsolata}

\usepackage{graphicx}
\usepackage{array}
\usepackage[hypcap=false]{caption}
\usepackage{enumitem}
\usepackage{amsmath}
\usepackage{algorithm}
\usepackage{algpseudocode}
\usepackage{cuted}
\usepackage{placeins}
\usepackage{xcolor}
\usepackage{tcolorbox}

\usepackage{graphicx}
\usepackage{booktabs}
\usepackage{multirow}
\usepackage[table]{xcolor}

\newif\ifgaramondteaser

\garamondteaserfalse    % Spectral
\ifgaramondteaser
    \usepackage{ebgaramond}
\else
    \usepackage{spectral}
\fi

\edef\teaserfamily{\rmdefault}
\renewcommand{\rmdefault}{ptm}

\newcommand{\teaserfont}{%
    \fontfamily{\teaserfamily}\selectfont
}

\newcolumntype{C}[1]{%
    >{\centering\arraybackslash}m{#1}%
}

\definecolor{nonsenseRed}{RGB}{155,58,8}

\newcommand{\teaserStrongRule}{%
    \noalign{\hrule height 0.75pt}
}

\newcommand{\QRPScoreLoss}[2]{%
    \textbf{Dense Similarity Score (#1):}\ #2%
}

\newcommand{\NonsenseScoreWin}[2]{%
    \textbf{Dense Similarity Score (#1):}\ %
    {\color{nonsenseRed}\textbf{#2}}%
}

\title{NONSENSE: Non-Semantic Strings Expose Semanticity Gaps in Dense Retrieval}

\author{Ankit Gupta\\
  Department of Computer Science and Engineering\\Michigan State University, East Lansing, United States \\
  \texttt{guptaa23@msu.edu}
}

\begin{document}
\maketitle

\begingroup
\renewcommand{\thefootnote}{}
\begin{NoHyper}
\footnotetext{\normalfont\footnotesize Accepted at EACL 2027 (Main Conference), Athens.}
\end{NoHyper}
\addtocounter{footnote}{-1}
\endgroup

\begin{abstract}
Dense retrieval is central to modern search, question answering, and retrieval-augmented generation, where high similarity is often treated as evidence of semantic relevance. We stress-test this assumption with NONSENSE, a black-box optimization framework that constructs visibly non-semantic strings by optimizing only their dense similarity to the query. Across seven retrieval datasets and five single-vector retrievers, these strings frequently outscore the retriever's own best judged query-relevant passage (QRP) and often rank first in the full corpus. The result indicates a semanticity gap in the evaluated dense retrievers: strings with no coherent passage-level semantic content can lie closer to the query in embedding space than genuine relevant evidence. These findings highlight the importance of evaluating dense retrievers not only by whether they retrieve relevant documents, but also by whether their similarity scores reliably reflect semantic evidence.
\end{abstract}
\vspace{-3.5mm}

\begin{figure*}[!t]
\centering
\vspace{-0.8mm}

\begingroup
\teaserfont
\small

\setlength{\tabcolsep}{8pt}
\setlength{\arrayrulewidth}{0.40pt}

\begin{tabular}{|C{0.47\textwidth}|C{0.47\textwidth}|}

% ============================================================
% HEADER
% ============================================================
\teaserStrongRule

% Main header:
% -1.6ex gives extra breathing room BELOW the header text.
\rule[-1.6ex]{0pt}{5.0ex}
{\large\bfseries Query-Relevant Passage (QRP)}
&
{\large\bfseries\color{nonsenseRed}
Generated NONSENSE String}
\\

\teaserStrongRule

% ============================================================
% SCIFACT QUERY
% ============================================================
\multicolumn{2}{|c|}{%
    \rule[-0.65ex]{0pt}{3.05ex}%
    \bfseries
    SciFact Query: Insomnia can be effectively treated with cognitive behavioral therapy.
}
\\

% ============================================================
% SCIFACT CONTENT
% ============================================================
\parbox[c][24.5mm][c]{0.96\linewidth}{%
    \centering

    A clinical approach to circadian rhythm sleep disorders..
    Circadian rhythm sleep disorders are characterized by
    complaints of insomnia...

    \vspace{2.8mm}

    {\footnotesize
    \QRPScoreLoss{E5}{0.854}}
}
&
\parbox[c][24.5mm][c]{0.96\linewidth}{%
    \centering

    {\footnotesize\ttfamily
    \begin{tabular}{@{}c@{}}
    -(S\_ \{|?0 =\&xP G\textbackslash{}V! eG.\textbackslash{} S0.D mj;l ((le\\[-0.6pt]
    (-?- !is\textbackslash{} s\textbackslash{}O, |!S\textbackslash{} o-b\textbackslash{} bs]\} ]++e |-e1\\[-0.6pt]
    -!*? a-\{- z?S] \&i\&5 9\&\%8 [i:A RnD/ |w\&d...
    \end{tabular}}

    \vspace{2.8mm}

    {\footnotesize
    \NonsenseScoreWin{E5}{0.914}}
}
\\

% ============================================================
% DATASET DIVIDER
% ============================================================
\teaserStrongRule

% ============================================================
% NFCORPUS QUERY
% ============================================================
\multicolumn{2}{|c|}{%
    \rule[-0.65ex]{0pt}{3.05ex}%
    \bfseries
    NFCorpus Query: Do Vegetarians Get Enough Protein?
}
\\

% ============================================================
% NFCORPUS CONTENT
% ============================================================
\parbox[c][24.5mm][c]{0.96\linewidth}{%
    \centering

    Current protein intake in America: analysis of the
    National Health and Nutrition Examination Survey,
    2003-2004.. In recent years there...

    \vspace{2.8mm}

    {\footnotesize
    \QRPScoreLoss{E5}{0.837}}
}
&
\parbox[c][24.5mm][c]{0.96\linewidth}{%
    \centering

    {\footnotesize\ttfamily
    \begin{tabular}{@{}c@{}}
    x\_|| L.0) q\&\#a .r/u -{}-\_i /Xql c\{IB ++\{\{\\[-0.6pt]
    -\{[\_=\&=] .;B) t4Vp +kgS :x1k );eW ;g*H\\[-0.6pt]
    2G,E EG,E ;e,B ,B;b [[:[ :\&[[ /-(? )[-(...
    \end{tabular}}

    \vspace{2.8mm}

    {\footnotesize
    \NonsenseScoreWin{E5}{0.880}}
}
\\

\teaserStrongRule

\end{tabular}
\endgroup

\vspace{0.2mm}

\caption{\textbf{Representative NONSENSE strings illustrating the semanticity gap.} Optimized NONSENSE strings frequently outscore the retriever's best judged query-relevant passage (QRP) under the same scoring model (E5), despite lacking any coherent semantic content. See Tables~\ref{tab:samples-e5}--\ref{tab:samples-contriever} for additional qualitative examples.}

\label{fig:teaser}

\vspace{-1mm}
\end{figure*}

\section{Introduction}

Dense retrieval is now standard infrastructure for open-domain question answering and retrieval-augmented generation~\citep{karpukhin2020dpr,lewis2020rag,petroni2021kilt}. A common operational assumption follows: if a candidate lies very close to a query in the retriever's embedding space, it should contain at least some meaningful evidence relevant to that query. This paper tests that assumption directly. We ask whether a dense retriever can place a visibly non-semantic string closer to the query than its best judged query-relevant passage. Throughout, we use \emph{semantic evidence} to mean readable passage content whose meaning bears on the query, and \emph{semanticity gap} to denote a separation between such evidence and proximity to the query in the retriever's embedding space.

\noindent
We introduce \textbf{NONSENSE}: \textbf{Non}-\textbf{S}emantic Strings Expose Semanticity Gaps in D\textbf{ense} Retrieval, a diagnostic probe for single-vector dense retrieval. Given a query and retriever, it initializes a fixed-length visible filler string and proposes one-character replacements. It accepts a replacement only when the candidate's dense similarity to the query improves, rejects candidates containing query terms, and finally inserts the optimized string into the full corpus for ranking. The output is not a prompt injection, does not rely on hidden characters, and does not repeat query terms; it is an overt character string constructed without coherent passage-level semantic content. Operationally, these strings contain no banned query terms or hidden characters, form essentially no ordinary dictionary words, exhibit atypical character composition, and receive little lexical support under BM25-style retrieval~\citep{robertson1994okapi}. Prior work on adversarial semantic collisions shows that semantically unrelated texts can nevertheless receive high model similarity~\citep{song2020semanticcollisions}; NONSENSE asks the stricter QRP-relative question of whether a standalone non-semantic artifact can outrank the retriever's own strongest judged evidence. Figure~\ref{fig:teaser} illustrates this failure mode.

\noindent
For each query, we compare the optimized NONSENSE score with the highest-scoring judged query-relevant passage (QRP) under the same retriever. Across seven datasets and five single-vector retrievers, NONSENSE Score exceeds QRP Score in 32 of 35 dataset--retriever pairs, with full-corpus median rank 1 in 28 of 35 pairs. ArguAna provides a recurring boundary, while ColBERTv2 late interaction~\citep{santhanam2022colbertv2} is more resistant overall, showing that the effect varies with both the model--dataset setting and the scoring architecture. Practically, these results motivate validating high-scoring retrievals with lexical or semantic checks before downstream use.

\noindent
In a nutshell, this paper makes the following contributions: (i) a black-box, fixed-length character-replacement probe that synthesizes standalone non-semantic strings using only retriever scores while forbidding query-term copying; (ii) a QRP-relative, full-corpus evaluation showing that such strings can outscore the retriever's strongest judged evidence and occupy the top of the embedding-space ranking across modern single-vector retrievers and diverse retrieval domains; and (iii) budget, lexical, control, ablation, transfer, and architectural analyses that rule out trivial lexical explanations, characterize model--dataset variation, and bound the effect across retrieval scoring architectures.

\section{Related Work}

Dense retrieval maps queries and documents into vector spaces for nearest-neighbor search, spanning dual-encoder retrieval, hard-negative training, distillation, and unsupervised contrastive learning~\citep{karpukhin2020dpr,xiong2021ance,qu2021rocketqa,hofstatter2021tasb,izacard2022contriever}. E5 and BGE-base are widely used embedding families~\citep{wang2022e5,xiao2024cpack}, while late-interaction and reranking models introduce explicit query--candidate interaction at additional computational cost~\citep{khattab2020colbert,nogueira2020monot5}. Sparse lexical retrieval such as BM25 remains a standard comparison point~\citep{robertson1994okapi}. Benchmarks such as MS MARCO, BEIR, and KILT primarily evaluate whether relevant documents are retrieved~\citep{nguyen2016msmarco,thakur2021beir,petroni2021kilt}. NONSENSE asks a complementary question: can a candidate with no coherent passage-level semantic content nevertheless become one of the retriever's nearest neighbors?

\noindent
Adversarial NLP has shown that character-, token-, and black-box methods can manipulate classifiers and entailment systems~\citep{ebrahimi2018hotflip,gao2018deepwordbug,wallace2019triggers,jin2020textfooler,li2020bertattack,morris2020textattack}, paralleling classic adversarial examples in vision~\citep{szegedy2014intriguing,goodfellow2015explaining}. Behavioral testing similarly motivates targeted probes beyond aggregate accuracy~\citep{ribeiro2020checklist}, and fooling images provide a useful analogy: inputs lacking human-recognizable content can nevertheless elicit highly confident model responses~\citep{nguyen2015fooling}. Closely related phenomena have also been studied directly in retrieval. Adversarial semantic collisions construct semantically unrelated texts that models judge as similar~\citep{song2020semanticcollisions}; dense-index analyses show that false-positive susceptibility can increase with index size~\citep{reimers2021curse}; corpus-poisoning methods optimize inserted passages to attract queries~\citep{zhong2023poisoning,zou2025poisonedrag}; and black-box dense-retrieval attacks perturb target documents to increase their retrievability~\citep{liu2023blackbox}. These lines of work show that closely related failure modes can arise across a range of retrieval settings, but with different inputs, objectives, assumptions, and success criteria than those considered by NONSENSE.

\noindent
NONSENSE isolates a different diagnostic object and criterion. It starts from visible filler and synthesizes a standalone, query-specific string using only retriever scores, with no source or target document and no semantic-preservation constraint. Query-term copying is explicitly forbidden, and success requires the resulting string, after insertion into the full corpus, to outscore the same retriever's highest-scoring judged query-relevant passage (QRP). The resulting artifact is not merely irrelevant to the query or adversarially modified from a document; it is constructed to contain no coherent passage-level semantic content. NONSENSE therefore directly tests whether nearest-neighbor status can be achieved without semantic evidence, relative to the retriever's own strongest judged evidence.

\section{Method}
Let $f_q$ and $f_p$ be the retriever's query and passage encoders. For a query $q$, NONSENSE maintains a fixed-length candidate string $s$, initialized by repeating ``\texttt{:::: }'' to 256 characters in the default setting. At each generation, the algorithm samples one mutable position and replaces its character with a visible character drawn from the active alphabet; with frozen spaces, positions initialized as spaces are never mutated. Query terms are lowercased alphanumeric spans matched by \texttt{[a-z0-9]+}; any candidate containing a query term of length at least three is rejected before scoring. A replacement is accepted only if it improves cosine similarity $\cos(f_q(q), f_p(s))$. Thus the main experiment is fixed-length and replacement-only. Algorithm~\ref{alg:nonsense} summarizes the complete optimization and evaluation procedure.

After $50,000$ proposal steps, we insert the final optimized string into the full corpus and rank it by dense similarity. Let $\mathcal{Q}$ denote the sampled query set, $s_q$ the optimized NONSENSE string for query $q$, and $\mathrm{QRel}(q)$ the set of judged relevant passages for $q$. We define the query-relevant passage (QRP) score as the highest dense similarity among these judged relevant passages:
\[
\mathrm{QRP}(q)
=
\max_{d \in \mathrm{QRel}(q)}
\mathrm{score}(q,d).
\]
The reported Success Rate is the empirical fraction of sampled queries for which NONSENSE exceeds this QRP score:
\[
\widehat{\mathrm{Success}}
=
\frac{1}{|\mathcal{Q}|}
\sum_{q \in \mathcal{Q}}
\mathbf{1}
\left[
\mathrm{score}(q,s_q) > \mathrm{QRP}(q)
\right].
\]

\paragraph{Metrics.}
\textbf{QRP Score} is the dense similarity of the highest-scoring judged query-relevant passage for a query, while \textbf{NONSENSE Score} is the similarity of the corresponding optimized string. \textbf{Success Rate} is the empirical fraction defined above and is reported as a percentage in Table~\ref{tab:main-results}. \textbf{Full-Corpus Median Rank} is the median rank of the inserted NONSENSE string after ranking it against the complete corpus. Because QRP is selected by the same retriever being evaluated, this comparison asks whether NONSENSE can outperform that model's strongest judged relevant evidence rather than an arbitrary relevant passage. Thus, a query counts as a success only when the optimized non-semantic string receives a higher score than the retriever's own highest-scoring judged relevant passage.

\begin{table*}[t]
\centering
\scriptsize
\setlength{\tabcolsep}{4.5pt}
\renewcommand{\arraystretch}{1.14}

\resizebox{\textwidth}{!}{%
\begin{tabular}{llccccccc}
\toprule
\textbf{Retriever} & \textbf{Metric}
& \shortstack{\textbf{SciFact}\\{\tiny 5,183 docs}}
& \shortstack{\textbf{SciDocs}\\{\tiny 25,657 docs}}
& \shortstack{\textbf{FIQA}\\{\tiny 57,600 docs}}
& \shortstack{\textbf{ArguAna}\\{\tiny 8,674 docs}}
& \shortstack{\textbf{NFCorpus}\\{\tiny 3,633 docs}}
& \shortstack{\textbf{CQA/Android}\\{\tiny 22,998 docs}}
& \shortstack{\textbf{CQA/English}\\{\tiny 40,221 docs}} \\
\midrule

& QRP Score
& 0.836 & 0.828 & 0.853 & 0.821 & 0.813 & 0.843 & 0.829 \\
& NONSENSE Score
& \textbf{0.902} & \textbf{0.892} & \textbf{0.894} & \textbf{0.879}
& \textbf{0.901} & \textbf{0.890} & \textbf{0.894} \\
\rowcolor{gray!7} \cellcolor{white}
& \quad Success Rate (\%)
& 97.7\% & 97.4\% & 91.7\% & 99.2\% & 97.1\% & 94.5\% & 97.7\% \\
\rowcolor{gray!7} \cellcolor{white}
\multirow{-4}{*}{\textbf{E5}}
& \quad Full-Corpus Median Rank
& 1 & 1 & 1 & 2 & 1 & 1 & 1 \\

\midrule

& QRP Score
& 0.800 & 0.789 & 0.784 & \textbf{0.812} & 0.721 & 0.772 & 0.763 \\
& NONSENSE Score
& \textbf{0.866} & \textbf{0.854} & \textbf{0.838} & 0.808
& \textbf{0.840} & \textbf{0.823} & \textbf{0.825} \\
\rowcolor{gray!7} \cellcolor{white}
& \quad Success Rate (\%)
& 85.2\% & 87.1\% & 80.7\% & 45.6\% & 92.8\% & 73.4\% & 71.1\% \\
\rowcolor{gray!7} \cellcolor{white}
\multirow{-4}{*}{\textbf{BGE-base}}
& \quad Full-Corpus Median Rank
& 1 & 1 & 1 & 4 & 1 & 1 & 1 \\

\midrule

& QRP Score
& 0.541 & 0.510 & 0.495 & 0.528 & 0.426 & 0.595 & 0.518 \\
& NONSENSE Score
& \textbf{0.721} & \textbf{0.646} & \textbf{0.628} & \textbf{0.588}
& \textbf{0.650} & \textbf{0.654} & \textbf{0.628} \\
\rowcolor{gray!7} \cellcolor{white}
& \quad Success Rate (\%)
& 95.6\% & 92.8\% & 90.6\% & 75.9\% & 99.0\% & 67.2\% & 77.3\% \\
\rowcolor{gray!7} \cellcolor{white}
\multirow{-4}{*}{\textbf{Contriever}}
& \quad Full-Corpus Median Rank
& 1 & 1 & 1 & 2 & 1 & 4 & 2 \\

\midrule

& QRP Score
& 0.622 & 0.584 & 0.639 & \textbf{0.670} & 0.488 & 0.677 & 0.614 \\
& NONSENSE Score
& \textbf{0.753} & \textbf{0.748} & \textbf{0.715} & 0.649
& \textbf{0.715} & \textbf{0.768} & \textbf{0.750} \\
\rowcolor{gray!7} \cellcolor{white}
& \quad Success Rate (\%)
& 84.0\% & 95.3\% & 68.8\% & 42.2\% & 92.6\% & 75.0\% & 76.6\% \\
\rowcolor{gray!7} \cellcolor{white}
\multirow{-4}{*}{\textbf{Qwen3-0.6B}}
& \quad Full-Corpus Median Rank
& 1 & 1 & 1 & 4 & 1 & 1 & 1 \\

\midrule

& QRP Score
& 0.617 & 0.600 & 0.645 & \textbf{0.759} & 0.533 & 0.654 & 0.581 \\
& NONSENSE Score
& \textbf{0.722} & \textbf{0.728} & \textbf{0.692} & 0.701
& \textbf{0.695} & \textbf{0.731} & \textbf{0.697} \\
\rowcolor{gray!7} \cellcolor{white}
& \quad Success Rate (\%)
& 82.4\% & 94.9\% & 67.6\% & 19.1\% & 94.1\% & 78.9\% & 77.3\% \\
\rowcolor{gray!7} \cellcolor{white}
\multirow{-4}{*}{\textbf{BGE-M3}}
& \quad Full-Corpus Median Rank
& 1 & 1 & 1 & 15 & 1 & 1 & 1 \\

\bottomrule
\end{tabular}%
}

\caption{\textbf{NONSENSE strings frequently outscore the retriever's highest-scoring judged query-relevant passage.} Across datasets and single-vector retrievers, optimized NONSENSE strings often achieve higher dense similarity than the highest-scoring query-relevant passage (QRP) and rank at or near the top of the full corpus. QRP Score and NONSENSE Score report their respective mean dense similarities. Indented rows report complementary evaluation metrics: Success Rate (\%) is the percentage of queries for which NONSENSE outperforms QRP, and Full-Corpus Median Rank is the median corpus rank of the inserted NONSENSE string. Corpus sizes are shown below each dataset name. Results use the same query batch size and optimization budget across settings. Bold marks the higher score between QRP and NONSENSE.}

\label{tab:main-results}
\end{table*}

\begin{figure*}[!t]
\centering
\includegraphics[width=1.0\textwidth]{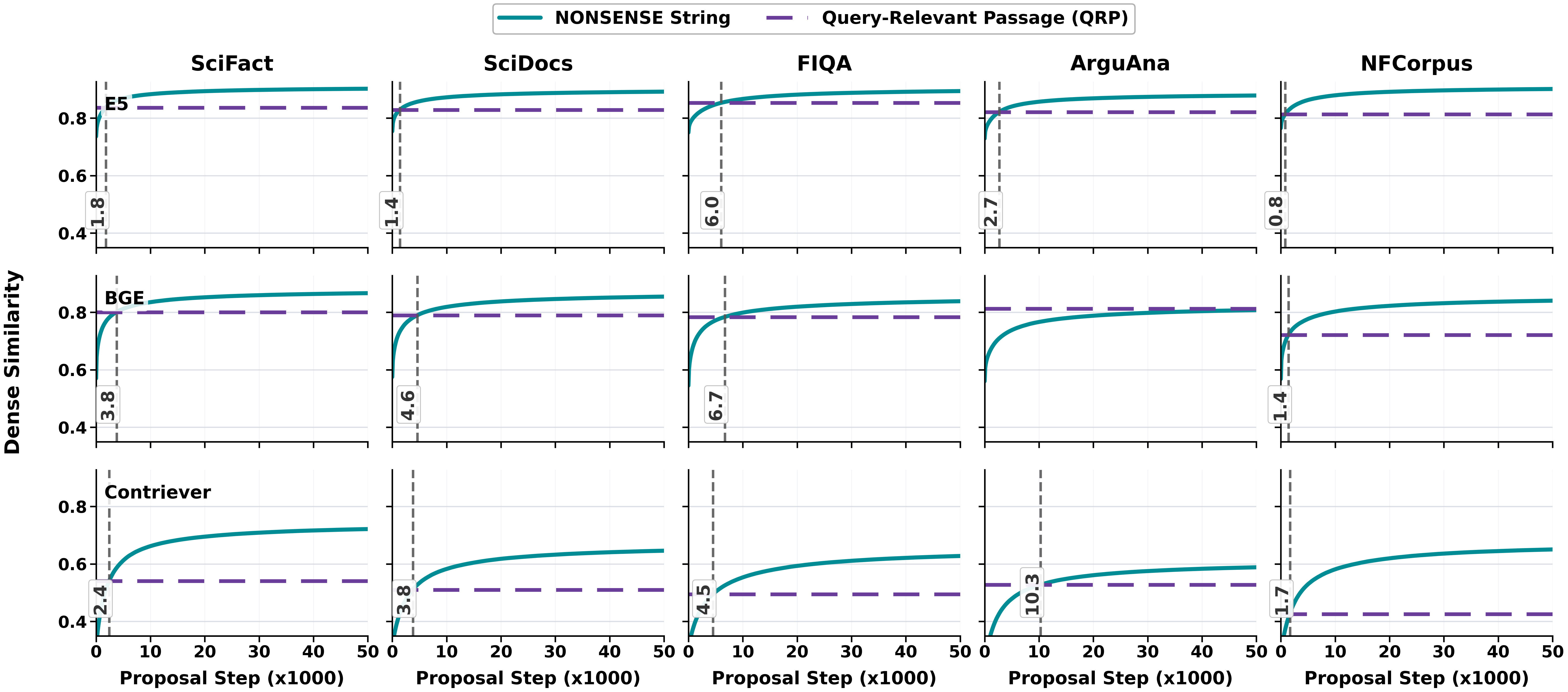}

\caption{\textbf{NONSENSE similarity rises rapidly and often overtakes query-relevant evidence early in optimization.} Solid curves show the mean NONSENSE score, while dashed lines show the corresponding mean query-relevant passage (QRP) score. Vertical dotted markers and boxed values indicate the first mean overtake point. These early crossings show that the semanticity gap emerges well before the final optimization budget.}

\label{fig:score-curves}
\end{figure*}

\section{Experiments and Results}

We evaluate five single-vector retrievers: E5-base-v2 (E5)~\citep{wang2022e5}, BGE-base-en-v1.5 (BGE-base)~\citep{xiao2024cpack}, Contriever~\citep{izacard2022contriever}, Qwen3-Embedding-0.6B (Qwen3-0.6B)~\citep{zhang2025qwen3embedding}, and BGE-M3~\citep{chen2024m3embedding}. We evaluate them on SciFact, SciDocs, FIQA, ArguAna, and NFCorpus~\citep{wadden2020scifact,cohan2020specter,maia2018fiqa,wachsmuth2018arguana,boteva2016nfcorpus}, together with the Android and English subsets of CQADupStack (CQA/Android and CQA/English)~\citep{hoogeveen2015cqadupstack}, using 256 sampled queries, full-corpus ranking, and $50{,}000$ proposal steps.

\paragraph{NONSENSE strings frequently outscore judged evidence.}
Table~\ref{tab:main-results} shows the central semanticity gap: optimized non-semantic strings frequently receive higher retrieval scores than the retriever's strongest judged query-relevant passage. NONSENSE Score exceeds QRP Score in 32 of the 35 dataset--retriever pairs, and the inserted NONSENSE string has full-corpus median rank 1 in 28 of 35 pairs. The pattern is broad across retrievers and extends to the community question-answering datasets. ArguAna provides a recurring boundary, with QRP remaining higher for BGE-base, Qwen3-0.6B, and BGE-M3. This variation shows that the effect is model--dataset dependent, while the strong performance across many other settings demonstrates that non-semantic strings can consistently outrank judged evidence under the evaluated scoring function.

\paragraph{The effect weakens under stronger query--candidate interaction.}
Appendix Table~\ref{tab:colbert-results} shows that ColBERTv2 late interaction~\citep{santhanam2022colbertv2} is more resistant: QRP scores higher on five of seven datasets, with NFCorpus and CQA/English as exceptions. This variation shows that the effect is not universal across retrieval architectures.

\paragraph{Optimization reaches the gap quickly.}
The detailed optimization analyses use E5, BGE-base, and Contriever on SciFact, SciDocs, FIQA, ArguAna, and NFCorpus. Figure~\ref{fig:score-curves} shows that NONSENSE often overtakes QRP within 1k--5k proposals, with NFCorpus/E5 crossing by step 800 and SciDocs/E5 by step 1,400. Scores continue rising afterward, suggesting the effect is not merely favorable initialization. ArguAna/BGE-base again provides the boundary: its curve approaches but does not cross QRP. Table~\ref{tab:budget-success} and Appendix Table~\ref{tab:margin-budget} show the same trend: overall QRP-relative success reaches 59.1\% by 5k and 72.9\% by 10k proposals, and the final median NONSENSE--QRP margin is positive for 14 of 15 dataset--retriever pairs.

\paragraph{Controls rule out lexical copying.}
Appendix Tables~\ref{tab:lexical-token} and~\ref{tab:cheap-ablations} separate the dense effect from trivial alternatives. Blank, random, and matched-random strings have near-zero success, while query-term repetition wins only by copying query terms and becoming a BM25 top hit. Optimized NONSENSE instead succeeds with zero query-term overlap and BM25@1 of zero in the SciFact control setting; across these runs, query-term rates are 0.0\%, BM25@1 is 1.9\% overall, dictionary-word counts are essentially zero, and most characters are non-alphanumeric. Thus, the strings are model-input artifacts rather than readable evidence or lexical copies.

\paragraph{Practical Implications.} A dense retriever ranking nonsensical nonsemantic artifacts above genuine relevant evidence represents a fundamental failure mode: inputs with no coherent meaning should not displace semantically relevant passages at the top of the ranking. This is more than an isolated scoring anomaly, because retrieval scores can determine which candidates are exposed to later stages of a system. If reproduced across multiple query-specific candidates, such artifacts could crowd useful evidence out of the highest-ranked results and distort the candidate set before any downstream reasoning occurs. If they survive filtering or reranking, they could then propagate into downstream RAG pipelines and reduce the quality or availability of retrieved evidence. These results therefore motivate treating dense similarity as a retrieval signal rather than sufficient evidence of semantic relevance, and complementing it with stronger candidate validation, reranking, or semantic verification.

\begin{table}[t]
\centering
\small
\setlength{\tabcolsep}{9pt}
\begin{tabular}{lrrrr}
\toprule
Retriever & 1k & 5k & 10k & 50k \\
\midrule
E5 & 28.8\% & 72.4\% & 85.4\% & 96.6\% \\
BGE-base & 16.2\% & 46.5\% & 59.1\% & 78.3\% \\
Contriever & 16.6\% & 58.4\% & 74.2\% & 90.8\% \\
Overall & 20.5\% & 59.1\% & 72.9\% & 88.6\% \\
\bottomrule
\end{tabular}

\caption{\textbf{Success Rate (\%) of NONSENSE at smaller optimization budgets.} Results aggregate SciFact, SciDocs, FIQA, ArguAna, and NFCorpus for each retriever. Entries report the percentage of query cases for which the snapshot NONSENSE score exceeds QRP after the stated number of proposal steps. The 50k column corresponds to the final optimization budget used in Table~\ref{tab:main-results}.}
\label{tab:budget-success}
\end{table}

\section{Conclusion}

NONSENSE shows that, in the evaluated single-vector dense retrievers, passage-like meaning and embedding-space proximity can separate sharply. QRP-relative success asks whether a non-passage can beat the strongest judged evidence under the same model, score, and corpus; frequently, it can. Across seven datasets and five single-vector retrievers, NONSENSE Score exceeds QRP Score in 32 of 35 dataset--retriever pairs, with full-corpus median rank 1 in 28 of 35 pairs. ColBERTv2 shows a weaker pattern, emphasizing that the effect depends on the scoring architecture rather than applying uniformly across retrieval systems. Because NONSENSE uses only black-box scoring, iterative optimization can synthesize non-semantic strings that score among a retriever's strongest matches. This is diagnostic but operationally relevant: high first-stage similarity should not be treated as sufficient semantic evidence.

\section{Limitations}

NONSENSE is an offline diagnostic evaluated on five single-vector retrievers across seven English retrieval datasets under exact scoring. We also evaluate ColBERTv2, but do not exhaust late-interaction or reranking architectures. Hybrid retrieval, approximate-nearest-neighbor indexing, multilingual settings, and end-to-end RAG remain outside scope. Cross-model transfer across E5, BGE-base, and Contriever is limited, suggesting that the optimized strings are largely model-specific. Finally, the strings are visibly non-semantic and not intended as human-facing evidence; the probe tests whether retrieval scoring alone can place such inputs above judged relevant passages.

\clearpage

\bibliography{custom}

\clearpage

\appendix

\section*{Appendix}

\noindent
This appendix provides additional experimental details and supporting analyses for the results in the main paper, including implementation and evaluation setup, control and lexical diagnostics, optimization-budget analyses, architectural scope, cross-model transfer, and qualitative examples.

\section{Detailed Experimental Setup}

\paragraph{Evaluation scope.}
The main single-vector evaluation spans five retrievers---E5-base-v2, BGE-base-en-v1.5, Contriever, Qwen3-Embedding-0.6B, and BGE-M3---across seven datasets: SciFact, SciDocs, FIQA, ArguAna, NFCorpus, CQA/Android, and CQA/English. Unless otherwise stated, all evaluations use 256 sampled queries, three seeds, and $50{,}000$ proposal steps per query.

The 256-query sample is drawn from the judged-query set for each dataset. This gives a fixed cross-dataset evaluation size while remaining below the available test-query count for every dataset in our study. SciFact has 300 test queries, NFCorpus 323, FIQA 648, SciDocs 1,000, ArguAna 1,406, CQA/Android 699, and CQA/English 1,570. The choice also aligns with the batched GPU execution used by the optimizer, which evaluates one proposal per query in batches of 256.

Additional budget, ablation, lexical, control, and transfer analyses are reported for E5, BGE-base, and Contriever on SciFact, SciDocs, FIQA, ArguAna, and NFCorpus. ColBERTv2 is evaluated separately as a late-interaction model on all seven datasets.

\paragraph{Retrieval setup.}
For the single-vector retrievers, we use cosine-normalized dense similarity. E5 uses the standard query and passage prefixes. For each dataset--retriever pair, we precompute full-corpus passage embeddings and rank the final NONSENSE string after inserting it into the complete corpus. The datasets cover scientific fact verification (SciFact), scientific document retrieval (SciDocs), financial question answering (FIQA), argument retrieval (ArguAna), medical information retrieval (NFCorpus), and community question answering (CQA/Android and CQA/English). Corpus sizes are listed in Table~\ref{tab:main-results}. We use judged relevant documents from the dataset qrels to define QRP. When a query has multiple judged relevant passages, QRP is the one with the highest retrieval score under the evaluated retriever.

\paragraph{Black-box optimization.}
The optimizer only queries the retriever as a scorer. It does not use gradients, model weights, token-level saliency, or training data. Each proposal replaces one visible character, keeping the string length fixed, and is accepted only when the retrieval score improves. The query-term ban is applied before scoring a candidate, so accepted strings cannot succeed by copying obvious lexical evidence from the query. In the main setting, the edit alphabet is ASCII letters, digits, punctuation characters \texttt{.,;:-\_/\textbackslash | \#@\%\&*+=?!()[]\{\}}, and a small set of visible non-ASCII symbols. Spaces appear in the initialization; with frozen spaces, those positions are not mutated. We do not use hidden control characters. The final reported rank is computed only after the optimized string is inserted into the full corpus.

\paragraph{Compute and runtime.}
Runs were executed in batched GPU jobs on H200 GPUs. Each run optimizes 256 queries jointly for $50{,}000$ proposal steps, proposing one replacement per query per step and encoding valid proposals in batches of 256. For E5, BGE-base, and Contriever runs on SciFact, SciDocs, FIQA, ArguAna, and NFCorpus, wall-clock time ranges from 106.3 to 189.5 minutes per dataset--retriever run, with a mean of 125.9 minutes and a median of 117.5 minutes. Each run optimizes 256 queries in parallel using only black-box scoring calls.

\begin{algorithm}[t]
\caption{NONSENSE}
\label{alg:nonsense}

\small
\algrenewcommand\algorithmicindent{0.85em}

\begin{algorithmic}[1]
\setlength{\itemsep}{0.10ex}

\Require query $q$, corpus $C$, retriever encoders $f_q,f_p$, length $L$, generations $T$, alphabet $\mathcal{A}$, frozen-space flag $z$
\State $e_q \gets f_q(q)$
\State $s \gets$ visible filler string of length $L$
\If{$z$}
    \State $\mathcal{M} \gets \{i \in \{1,\ldots,L\}: s_i \neq \texttt{space}\}$
\Else
    \State $\mathcal{M} \gets \{1,\ldots,L\}$
\EndIf
\State $\mathcal{B} \gets$ lowercased alphanumeric query terms of length $\geq 3$
\State $e_s \gets f_p(s)$; \quad $b \gets \cos(e_q,e_s)$
\For{$t=1,\ldots,T$}
    \State $v \gets \mathrm{false}$
    \For{$r=1,\ldots,50$}
        \State sample position $i \sim \mathrm{Uniform}(\mathcal{M})$ and character $a \sim \mathrm{Uniform}(\mathcal{A})$
        \If{$a=s_i$} \State \textbf{continue} \EndIf
        \State $s' \gets s$ with $s'_i \gets a$
        \If{$s'$ contains any term in $\mathcal{B}$} \State \textbf{continue} \EndIf
        \State $v \gets \mathrm{true}$
        \State \textbf{break}
    \EndFor
    \If{$v$}
        \State $e' \gets f_p(s')$; \quad $b' \gets \cos(e_q,e')$
        \If{$b' > b$}
            \State $s \gets s'$; \quad $e_s \gets e'$; \quad $b \gets b'$
        \EndIf
    \EndIf
\EndFor
\State insert $s$ into $C$ and rank it by dense similarity to $q$
\State compare $b$ to QRP score and record success and corpus rank

\end{algorithmic}
\end{algorithm}

\section{Interpreting the Control Baselines}

The controls are designed to rule out three simple explanations. First, the blank and random-string rows test whether arbitrary off-manifold text is already highly ranked. They are not: success is essentially zero and mean ranks are in the thousands. Second, matched random edits test whether the large edit budget alone creates strong candidates. It does not. Third, query-term repetition verifies that a lexical copying baseline can win, while also showing why it is a different phenomenon: it copies query terms and becomes a BM25 top hit. Optimized NONSENSE provides the diagnostic contrast: it wins dense similarity without query-term overlap and without lexical retrieval success.

\section{Budget and Lexical Diagnostics}

For E5, BGE-base, and Contriever on SciFact, SciDocs, FIQA, ArguAna, and NFCorpus, Table~\ref{tab:budget-success} reports QRP-relative success at earlier optimization budgets. The effect is visible well before the final $50{,}000$ proposals: aggregate success across these runs is 20.5\% at 1k, 59.1\% at 5k, and 72.9\% at 10k. This shows that the result does not depend solely on the final optimization endpoint. Table~\ref{tab:ablations} reports complementary SciFact ablations over initialization, length, space mutability, and edit alphabet.

Table~\ref{tab:lexical-token} audits lexical leakage in the final optimized strings. The query-term checks show no leakage, while BM25@1 remains low outside SciFact. We therefore treat BM25 as a diagnostic rather than part of the main dense-retrieval claim.

\section{Tokenization and Lexical Audit Details}

Table~\ref{tab:lexical-token} is designed to distinguish dense-similarity success from readable evidence, lexical copying, and ordinary sparse retrieval. Query-term rates are computed after lowercasing and extracting alphanumeric spans with the same rule used by the optimizer. Exact Rate requires a banned query term to appear as a complete alphanumeric span, while Query-Term Rate also counts substring matches. Dictionary words are counted over alphabetic spans after lowercasing; the near-zero counts indicate that optimized strings rarely form ordinary English words. Alpha Tokens count alphabetic spans regardless of whether they are dictionary words, and Non-Alnum. reports the fraction of characters outside letters and digits.

These diagnostics are lexical rather than mechanistic. They do not identify which tokenizer units, embedding axes, or pooling behaviors produce high dense similarity. Instead, they establish the narrower claim needed for the main result: optimized NONSENSE strings beat QRP without obvious query copying, BM25 success, or readable word formation. A fuller mechanistic account of tokenization patterns and embedding directions is left for future work.

\section{Architectural Scope}

Table~\ref{tab:colbert-results} evaluates NONSENSE with ColBERTv2 late interaction~\citep{santhanam2022colbertv2}. QRP receives the higher mean score on five of the seven datasets, while NFCorpus and CQA/English are exceptions, with QRP-relative success rates of 59.8\% and 60.9\%, respectively. As a separate pairwise check, a MiniLM-based cross-encoder~\citep{wang2020minilm} prefers QRP in 94.4\% of sampled QRP--NONSENSE pairs. Because the MiniLM experiment is a pairwise comparison rather than full-corpus retrieval, we treat it as supporting evidence about architectural scope rather than as an evaluation of a complete cross-encoder retrieval pipeline.

Together, these results show that the observed behavior varies across retrieval architectures. We treat this comparison as descriptive rather than causal: architecture, tokenization, score distributions, and dataset-specific QRP margins vary jointly.

\section{Additional Scope Notes}

The effect also varies substantially by model and dataset within the single-vector setting, with ArguAna providing a recurring boundary for several retrievers. This variation suggests that the probe can distinguish model--dataset pairs where non-semantic strings are easier or harder to place near queries. Future audits should test larger and multilingual corpora, approximate-nearest-neighbor retrieval stacks, hybrid systems, additional late-interaction and cross-encoder models, semantic-verifier defenses, and mechanistic analyses of tokenizer and embedding-space behavior.

\section{Diagnostic Interpretation}

The results should be read as evidence about the geometry exposed by the evaluated scoring functions. QRP-relative success is intentionally stringent because QRP is selected by the same retriever being audited: the comparison asks whether the model's own strongest judged evidence can be displaced by a non-passage under identical scoring. The lexical diagnostics then separate this effect from ordinary term matching. Taken together, the main table, score curves, and controls show that single-vector dense similarity can assign high rank to optimized non-passage inputs that fall outside the document assumptions under which retrieval outputs are normally consumed. The ColBERTv2 and MiniLM results further bound this interpretation by showing that the same behavior need not appear to the same degree under scoring architectures with explicit query--candidate interaction.

Table~\ref{tab:transfer} reports cross-model transfer diagnostics for E5, BGE-base, and Contriever. Strings optimized for a retriever usually remain strongest under that same retriever, while off-diagonal transfer is limited on SciFact, SciDocs, FIQA, and ArguAna. NFCorpus is the main exception: strings optimized for one model sometimes exceed another model's QRP and can land within the first few hundred target-corpus ranks. This pattern suggests that NONSENSE is not merely a universal trigger shared by all encoders; much of the effect is model-specific, with some dataset-dependent shared behavior.

Tables~\ref{tab:samples-e5}--\ref{tab:samples-contriever} provide representative qualitative examples, showing one randomly selected query per dataset with the target retriever's highest-scoring QRP, generated NONSENSE string, and dense similarity scores.

\begin{table}[t]
\centering
\footnotesize
\setlength{\tabcolsep}{4.8pt}
\renewcommand{\arraystretch}{1.8}

\begin{tabular}{lcccc}
\toprule
\textbf{Retriever}
& \shortstack{\textbf{QRP}\\\textbf{Score}}
& \shortstack{\textbf{NONSENSE}\\\textbf{Score}}
& \shortstack{\textbf{Success}\\\textbf{Rate (\%)}}
& \shortstack{\textbf{Median}\\\textbf{Rank}} \\
\midrule

\multicolumn{5}{l}{\textbf{Default}} \\[-1pt]
E5         & 0.838 & \textbf{0.902} & 98.4\% & 1 \\
BGE-base   & 0.802 & \textbf{0.865} & 85.2\% & 1 \\
Contriever & 0.540 & \textbf{0.718} & 95.3\% & 1 \\
\addlinespace[6pt]

\multicolumn{5}{l}{\textbf{Mutable spaces}} \\[-1pt]
E5         & 0.838 & \textbf{0.911} & 99.6\% & 1 \\
BGE-base   & 0.802 & \textbf{0.880} & 89.5\% & 1 \\
Contriever & 0.540 & \textbf{0.745} & 97.7\% & 1 \\
\addlinespace[6pt]

\multicolumn{5}{l}{\textbf{Length 128}} \\[-1pt]
E5         & 0.838 & \textbf{0.904} & 98.0\% & 1 \\
BGE-base   & 0.802 & \textbf{0.863} & 82.4\% & 1 \\
Contriever & 0.540 & \textbf{0.733} & 95.7\% & 1 \\
\addlinespace[6pt]

\multicolumn{5}{l}{\textbf{Length 512}} \\[-1pt]
E5         & 0.838 & \textbf{0.896} & 95.7\% & 1 \\
BGE-base   & 0.802 & \textbf{0.857} & 80.1\% & 1 \\
Contriever & 0.540 & \textbf{0.691} & 93.0\% & 1 \\
\addlinespace[6pt]

\multicolumn{5}{l}{\textbf{Init \texttt{aaaa}}} \\[-1pt]
E5         & 0.838 & \textbf{0.916} & 99.2\% & 1 \\
BGE-base   & 0.802 & \textbf{0.891} & 91.4\% & 1 \\
Contriever & 0.540 & \textbf{0.747} & 98.4\% & 1 \\
\addlinespace[6pt]

\multicolumn{5}{l}{\textbf{Init random}} \\[-1pt]
E5         & 0.838 & \textbf{0.912} & 99.6\% & 1 \\
BGE-base   & 0.802 & \textbf{0.882} & 89.8\% & 1 \\
Contriever & 0.540 & \textbf{0.738} & 96.5\% & 1 \\
\addlinespace[6pt]

\multicolumn{5}{l}{\textbf{No letters}} \\[-1pt]
E5         & \textbf{0.838} & 0.835 & 44.1\% & 2 \\
BGE-base   & \textbf{0.802} & 0.785 & 37.9\% & 2 \\
Contriever & 0.540 & \textbf{0.555} & 56.6\% & 2 \\

\bottomrule
\end{tabular}

\caption{\textbf{Ablation results on SciFact.}
Each ablation changes one component of the default NONSENSE configuration: Mutable spaces allows initialized space positions to mutate; Length changes the candidate-string length; Init \texttt{aaaa} and Init random change the initialization; and No letters restricts the edit alphabet to digits and symbols. Ablations use a single fixed run for controlled comparison; all other settings follow the default evaluation setup. Bold marks the higher score between QRP and NONSENSE.}

\label{tab:ablations}
\end{table}

\begin{table*}[t]
\centering
\small
\renewcommand{\arraystretch}{1.12}

\begin{tabular*}{\textwidth}{
@{\extracolsep{\fill}}
lcccccc
@{}
}
\toprule
\textbf{Retriever}
& \shortstack{\textbf{Query-Term}\\\textbf{Rate (\%)}}
& \shortstack{\textbf{Exact-Term}\\\textbf{Rate (\%)}}
& \shortstack{\textbf{BM25@1}\\\textbf{(\%)}}
& \shortstack{\textbf{Average Alphabetic}\\\textbf{Tokens}}
& \shortstack{\textbf{Average Dictionary}\\\textbf{Words}}
& \shortstack{\textbf{Non-Alnum.}\\\textbf{Characters (\%)}} \\
\midrule

E5
& 0.0\% & 0.0\% & 2.2\% & 50.9 & 0.00 & 62.1\% \\

BGE-base
& 0.0\% & 0.0\% & 1.8\% & 53.6 & 0.01 & 62.4\% \\

Contriever
& 0.0\% & 0.0\% & 1.8\% & 55.8 & 0.00 & 51.0\% \\

\midrule
\textbf{Overall}
& 0.0\% & 0.0\% & 1.9\% & 53.6 & 0.00 & 58.2\% \\

\bottomrule
\end{tabular*}

\caption{\textbf{Lexical diagnostics of optimized NONSENSE strings.}
Query-Term Rate measures whether any banned query term appears as a substring, while Exact-Term Rate requires an exact alphanumeric-span match. BM25@1 is the percentage of strings ranked first by BM25. Average Alphabetic Tokens and Average Dictionary Words report the number of alphabetic spans and common dictionary words per string, respectively; Non-Alphanumeric Characters is the percentage of characters that are not letters or digits.}

\label{tab:lexical-token}
\end{table*}

\begin{table*}[t]
\centering
\small
\renewcommand{\arraystretch}{1.10}

\begin{tabular*}{\textwidth}{
@{\extracolsep{\fill}}
llccccc
@{}
}
\toprule
\textbf{Retriever}
& \textbf{Metric}
& \textbf{SciFact}
& \textbf{SciDocs}
& \textbf{FIQA}
& \textbf{ArguAna}
& \textbf{NFCorpus} \\
\midrule

& Median Margin
& 0.065 & 0.064 & 0.040 & 0.057 & 0.082 \\
& IQR
& [0.039, 0.090]
& [0.039, 0.087]
& [0.021, 0.060]
& [0.041, 0.073]
& [0.051, 0.120] \\
\rowcolor{gray!7} \cellcolor{white}
\multirow{-3}{*}{\textbf{E5}}
& \shortstack[l]{Mean-Curve\\Overtake Step}
& 1,800 & 1,400 & 6,000 & 2,700 & 800 \\

\midrule

& Median Margin
& 0.061 & 0.065 & 0.057 & -0.006 & 0.109 \\
& IQR
& [0.017, 0.114]
& [0.026, 0.107]
& [0.012, 0.095]
& [-0.037, 0.029]
& [0.055, 0.173] \\
\rowcolor{gray!7} \cellcolor{white}
\multirow{-3}{*}{\textbf{BGE-base}}
& \shortstack[l]{Mean-Curve\\Overtake Step}
& 3,800 & 4,600 & 6,700 & Not crossed & 1,400 \\

\midrule

& Median Margin
& 0.176 & 0.132 & 0.135 & 0.061 & 0.209 \\
& IQR
& [0.090, 0.263]
& [0.069, 0.203]
& [0.067, 0.199]
& [0.004, 0.121]
& [0.131, 0.296] \\
\rowcolor{gray!7} \cellcolor{white}
\multirow{-3}{*}{\textbf{Contriever}}
& \shortstack[l]{Mean-Curve\\Overtake Step}
& 2,400 & 3,800 & 4,500 & 10,300 & 1,700 \\

\bottomrule
\end{tabular*}

\caption{\textbf{Score-margin and optimization-budget diagnostics.}
Margin is NONSENSE Score minus QRP Score for each query; Median Margin and IQR summarize the resulting query cases. Mean-Curve Overtake Step is the first proposal step at which the mean NONSENSE score in Figure~\ref{fig:score-curves} exceeds the mean QRP score. ArguAna/BGE-base is the only setting whose mean curve does not cross QRP.}

\label{tab:margin-budget}
\end{table*}

\begin{table*}[t]
\centering
\small
\setlength{\tabcolsep}{4.5pt}
\renewcommand{\arraystretch}{1.10}

\resizebox{\textwidth}{!}{%
\begin{tabular}{llcccccc}
\toprule
\textbf{Retriever}
& \textbf{Candidate}
& \shortstack{\textbf{QRP}\\\textbf{Score}}
& \shortstack{\textbf{Candidate}\\\textbf{Score}}
& \shortstack{\textbf{Success Rate}\\\textbf{(\%)}}
& \shortstack{\textbf{Mean Corpus}\\\textbf{Rank}}
& \shortstack{\textbf{Query-Term}\\\textbf{Rate (\%)}}
& \shortstack{\textbf{BM25@1}\\\textbf{(\%)}} \\
\midrule

\multirow{5}{*}{\textbf{E5}}
& \textbf{Optimized NONSENSE}
& 0.838 & \textbf{0.902} & 98.4\% & 1.0 & 0.0\% & 0.0\% \\
& Blank initialization
& \textbf{0.838} & 0.737 & 0.4\% & 1,802.6 & 0.0\% & 0.0\% \\
& Random string
& \textbf{0.838} & 0.742 & 0.4\% & 1,481.5 & 0.0\% & 0.0\% \\
& Matched random edits
& \textbf{0.838} & 0.743 & 0.8\% & 1,436.0 & 0.0\% & 0.0\% \\
& Query-term repetition
& 0.838 & \textbf{0.903} & 96.9\% & 1.0 & 100.0\% & 98.8\% \\

\midrule

\multirow{5}{*}{\textbf{BGE-base}}
& \textbf{Optimized NONSENSE}
& 0.802 & \textbf{0.865} & 85.2\% & 1.2 & 0.0\% & 0.0\% \\
& Blank initialization
& \textbf{0.802} & 0.571 & 0.0\% & 3,015.0 & 0.0\% & 0.0\% \\
& Random string
& \textbf{0.802} & 0.575 & 0.0\% & 2,819.9 & 0.4\% & 0.0\% \\
& Matched random edits
& \textbf{0.802} & 0.575 & 0.0\% & 2,803.7 & 0.0\% & 0.0\% \\
& Query-term repetition
& 0.802 & \textbf{0.915} & 97.7\% & 1.0 & 100.0\% & 98.8\% \\

\midrule

\multirow{5}{*}{\textbf{Contriever}}
& \textbf{Optimized NONSENSE}
& 0.540 & \textbf{0.718} & 95.3\% & 1.1 & 0.0\% & 0.0\% \\
& Blank initialization
& \textbf{0.540} & 0.272 & 1.6\% & 3,229.5 & 0.0\% & 0.0\% \\
& Random string
& \textbf{0.540} & 0.263 & 0.0\% & 3,496.7 & 0.0\% & 0.0\% \\
& Matched random edits
& \textbf{0.540} & 0.265 & 0.0\% & 3,472.0 & 0.0\% & 0.0\% \\
& Query-term repetition
& 0.540 & \textbf{0.818} & 99.2\% & 1.0 & 100.0\% & 98.8\% \\

\bottomrule
\end{tabular}%
}

\caption{\textbf{Control-baseline comparison on SciFact.}
Optimized NONSENSE is the default search result; Blank initialization is the unoptimized filler string; Random string samples a same-length random candidate; Matched random edits applies random edits under the same edit budget; and Query-term repetition fills the candidate with repeated query terms. Mean Corpus Rank is computed after inserting the candidate into the full corpus. Controls use a single fixed run for comparison, with all other settings following the default evaluation setup. Bold marks the higher score between QRP and the candidate.}

\label{tab:cheap-ablations}
\end{table*}

\begin{table*}[t]
\centering
\small
\setlength{\tabcolsep}{8pt}
\renewcommand{\arraystretch}{1.08}

\begin{tabular}{lcccc}
\toprule
\textbf{Dataset}
& \textbf{QRP Score}
& \textbf{NONSENSE Score}
& \textbf{Success Rate (\%)}
& \textbf{Full-Corpus Median Rank} \\
\midrule

SciFact
& \textbf{0.637} & 0.580 & 31.6\% & 5 \\

SciDocs
& \textbf{0.568} & 0.552 & 42.2\% & 16.5 \\

FIQA
& \textbf{0.580} & 0.543 & 39.5\% & 17 \\

ArguAna
& \textbf{0.687} & 0.533 & 2.3\% & 624.5 \\

NFCorpus
& 0.509 & \textbf{0.582} & 59.8\% & 2 \\

CQA/Android
& \textbf{0.623} & 0.572 & 32.0\% & 15 \\

CQA/English
& 0.598 & \textbf{0.648} & 60.9\% & 2 \\

\bottomrule
\end{tabular}

\caption{\textbf{ColBERTv2 late-interaction results.} Success Rate is the percentage of queries for which NONSENSE outperforms QRP; Full-Corpus Median Rank is the median corpus rank of the inserted NONSENSE string. Bold marks the higher score between QRP and NONSENSE.}

\label{tab:colbert-results}
\end{table*}

\begin{table*}[t]
\centering
\small
\setlength{\tabcolsep}{5.5pt}
\renewcommand{\arraystretch}{1.10}

\begin{tabular*}{\textwidth}{
@{\extracolsep{\fill}}
llcccccc
@{}
}
\toprule
\textbf{Dataset}
& \textbf{Source Retriever}
& \multicolumn{3}{c}{\textbf{Target-QRP Success Rate (\%)}}
& \multicolumn{3}{c}{\textbf{Mean Target Corpus Rank}} \\
\cmidrule(lr){3-5}
\cmidrule(lr){6-8}
&
& \textbf{E5}
& \textbf{BGE-base}
& \textbf{Contriever}
& \textbf{E5}
& \textbf{BGE-base}
& \textbf{Contriever} \\
\midrule

\multirow{3}{*}{SciFact}
& E5
& 98.4\% & 8.6\% & 5.1\%
& 1.0 & 190.2 & 1,207.1 \\
& BGE-base
& 8.6\% & 85.2\% & 3.5\%
& 179.4 & 1.2 & 1,862.2 \\
& Contriever
& 15.6\% & 6.6\% & 95.3\%
& 60.1 & 278.0 & 1.1 \\

\midrule

\multirow{3}{*}{SciDocs}
& E5
& 96.9\% & 1.2\% & 0.8\%
& 1.1 & 1,746.7 & 13,599.2 \\
& BGE-base
& 9.0\% & 86.7\% & 0.4\%
& 1,371.4 & 2.5 & 14,780.8 \\
& Contriever
& 12.5\% & 1.2\% & 92.2\%
& 1,233.7 & 3,192.2 & 2.2 \\

\midrule

\multirow{3}{*}{FIQA}
& E5
& 89.8\% & 0.8\% & 0.8\%
& 1.5 & 7,600.7 & 29,363.8 \\
& BGE-base
& 1.2\% & 79.7\% & 0.4\%
& 8,355.3 & 2.3 & 37,112.7 \\
& Contriever
& 4.3\% & 2.7\% & 89.5\%
& 4,432.8 & 9,156.9 & 3.6 \\

\midrule

\multirow{3}{*}{ArguAna}
& E5
& 98.8\% & 0.8\% & 0.0\%
& 2.0 & 2,568.3 & 5,339.4 \\
& BGE-base
& 2.0\% & 44.1\% & 0.0\%
& 1,162.6 & 11.5 & 6,400.1 \\
& Contriever
& 1.2\% & 0.4\% & 76.6\%
& 1,008.3 & 2,330.4 & 3.3 \\

\midrule

\multirow{3}{*}{NFCorpus}
& E5
& 96.9\% & 22.3\% & 12.5\%
& 1.1 & 288.5 & 1,377.8 \\
& BGE-base
& 32.0\% & 92.6\% & 11.3\%
& 313.6 & 1.3 & 1,721.2 \\
& Contriever
& 35.2\% & 19.9\% & 98.8\%
& 96.6 & 434.2 & 1.0 \\

\bottomrule
\end{tabular*}

\caption{\textbf{Cross-model transfer of optimized NONSENSE strings.}
Rows indicate the source retriever used for optimization, while columns indicate the target retriever used only for evaluation. Target-QRP Success Rate is the percentage of transferred strings whose target-retriever score exceeds that retriever's QRP score for the same query; Mean Target Corpus Rank is computed after inserting the transferred string into the target corpus. Diagonal entries correspond to same-model evaluation, while off-diagonal entries measure cross-retriever transfer. Transfer results use a single fixed run for controlled comparison.}

\label{tab:transfer}
\end{table*}

% ============================================================
% QUALITATIVE SAMPLE TABLES
% E5 / BGE / Contriever
%
% Uses existing:
%   \teaserfont
%   \definecolor{nonsenseRed}{RGB}{155,58,8}
%   array package
% ============================================================

% ------------------------------------------------------------
% Local helpers for qualitative tables
% ------------------------------------------------------------

% Full-width query row; wraps safely.
\newcommand{\QualQuery}[1]{%
    \multicolumn{2}{|
        >{\centering\arraybackslash}m{0.94\textwidth}|
    }{%
        \rule[-0.45ex]{0pt}{2.85ex}%
        \bfseries #1%
    }\\
}

% Left/QRP cell.
% #1 = exact passage
% #2 = model name
% #3 = score
\newcommand{\QualQRP}[3]{%
    \parbox[c][22mm][c]{0.96\linewidth}{%
        \centering
        #1\par
        \vfill
        {\footnotesize
        \textbf{Dense Similarity Score (#2):} #3}%
    }%
}

% Right/NONSENSE cell.
% #1 = exact generated string
% #2 = model name
% #3 = score
\newcommand{\QualNonsense}[3]{%
    \parbox[c][22mm][c]{0.96\linewidth}{%
        \centering
        {\footnotesize\ttfamily
        \sloppy
        #1\par}%
        \vfill
        {\footnotesize
        \textbf{Dense Similarity Score (#2):}
        {\color{nonsenseRed}\textbf{#3}}}%
    }%
}

% Strong horizontal partition
\newcommand{\QualRule}{%
    \noalign{\hrule height 0.75pt}
}

% ============================================================
% E5
% ============================================================

\begin{table*}[t]
\centering

\begingroup
\teaserfont
\footnotesize

\setlength{\tabcolsep}{7pt}
\setlength{\arrayrulewidth}{0.40pt}

\begin{tabular}{|
    >{\centering\arraybackslash}m{0.47\textwidth}|
    >{\centering\arraybackslash}m{0.47\textwidth}|
}

\QualRule

% Header
\rule[-0.8ex]{0pt}{4.0ex}
{\large\bfseries Query-Relevant Passage (QRP)}
&
{\large\bfseries\color{nonsenseRed}
Generated NONSENSE String}
\\

\QualRule

% ------------------------------------------------------------
% SciFact
% ------------------------------------------------------------
\QualQuery{SciFact Query: Activator-inhibitor pairs are provided dorsally by Admpchordin.}

\QualQRP{
Regulation of ADMP and BMP2/4/7 at Opposite Embryonic Poles Generates a Self-Regulating Morphogenetic Field. Embryos have the ability to self-regu ...
}{E5}{0.819}
&
\QualNonsense{
\_\%;; R/sn :X-N /tdi \%-!$\Delta$ 1mG) ChR3 IDL2 /PKN -n1( adFI );\%b *y(a )-!\textbackslash{} X;B- \%(-- S\%nQ $\times$\%-R (|\textbackslash{}s |-yY -\&A| ]:dR Z|\textbackslash{}R |-$\Delta$\_ N*c; ((c| -A-- n ...
}{E5}{0.910}
\\

\QualRule

% ------------------------------------------------------------
% SciDocs
% ------------------------------------------------------------
\QualQuery{SciDocs Query: Subjectivity and Sentiment Analysis of Modern Standard Arabic}

\QualQRP{
Sentiment Analyzer: Extracting Sentiments about a Given Topic using Natural Language Processing Techniques. We present Sentiment Analyzer (SA) tha ...
}{E5}{0.835}
&
\QualNonsense{
;.:. .\&.. /s:: O\textbackslash{}AY o()u :AlO /2]) -wAQ \textbackslash{}m\&O *YA: :E\&9 \%d83 \textbackslash{}I*i .:.: :M\_\textbackslash{} ay=: :\textbackslash{}i- *]++ *?\textbackslash{}; \textbackslash{}A/C (-+) \&T() *s(M \_\textbackslash{}a* ?\textbackslash{}R\textbackslash{} XYY, \textbackslash{}S(@ : ...
}{E5}{0.899}
\\

\QualRule

% ------------------------------------------------------------
% FIQA
% ------------------------------------------------------------
\QualQuery{FIQA Query: Something looks off about Mitsubishi financial data}

\QualQRP{
All but certainly, Mitsubishi is selling so cheaply because of the fuel scandal. It has been providing false fuel efficiency data for decades. As ...
}{E5}{0.822}
&
\QualNonsense{
\&\&1Y Yr)M t:(? !!!? :MxL Ie\_I (A:0 P9.8 (jDC r1\_R fxmR R$\Delta$s2 ,\}], -\textbackslash{}\_A *A*] ==-2 ,M5U -4:3 ]D:0 P$\times$6H -m.A mzR) 0PIX -9.5 ]\%=\& t?\&\% (5$\Delta$) M ...
}{E5}{0.897}
\\

\QualRule

% ------------------------------------------------------------
% ArguAna
% ------------------------------------------------------------
\QualQuery{ArguAna Query: Syria clearly meets the standards for intervention The Assad regime has clearly lost its legitimacy and has pre ...}

\QualQRP{
global middle east house would arm syrian rebels. The if the rebels are armed and the regime gets close to losing surely it will simply increase t ...
}{E5}{0.801}
&
\QualNonsense{
:n$\times$[ *\%\%0 (/qz \#\#:+ :*-a HYIr $\times$:QA :A(\} 0.3[ :*x: \&\%j\textbackslash{} s,Un /KD\textbackslash{} /nVJ :+60 M\#.3 *a\_n n\textbackslash{}a: $\times$*iW \_$\times$|$\times$ /WV3 \_X4\textbackslash{} \textbackslash{}x2\# $\times$]a: R*HN R/Yn -7,n $\times$ ...
}{E5}{0.890}
\\

\QualRule

% ------------------------------------------------------------
% NFCorpus
% ------------------------------------------------------------
\QualQuery{NFCorpus Query: Sexually Transmitted Fish Toxin}

\QualQRP{
Ciguatera fish poisoning. A southern California epidemic.. Ciguatera fish poisoning results from the bioconcentration of a variety of toxins produ ...
}{E5}{0.831}
&
\QualNonsense{
:=0\& u\_): SxFW !\_!! 2P\textbackslash{}L LH:: n,\%\# n888 -bJw NTR5 \textbackslash{}!0: \&yEM \_[[S =i\{\% 2\textbackslash{}7\textbackslash{} I\textbackslash{}hH *\textbackslash{}T\textbackslash{} s?7\textbackslash{} t8*\% 5[(s [Sx\# \#fDx \textbackslash{}4!7 /1-\textbackslash{} 8[7[ 5,7? mg\textbackslash{}Q K ...
}{E5}{0.903}
\\

\QualRule

\end{tabular}
\endgroup

\vspace{-0.6mm}

% Table 8 — E5

\caption{Representative E5 examples across five datasets. Each example shows a query, the retriever's highest-scoring judged query-relevant passage (QRP), and the optimized NONSENSE string, together with their dense similarity scores. Text is truncated for display.}
\label{tab:samples-e5}

\end{table*}

% ============================================================
% BGE
% ============================================================

\begin{table*}[t]
\centering

\begingroup
\teaserfont
\footnotesize

\setlength{\tabcolsep}{7pt}
\setlength{\arrayrulewidth}{0.40pt}

\begin{tabular}{|
    >{\centering\arraybackslash}m{0.47\textwidth}|
    >{\centering\arraybackslash}m{0.47\textwidth}|
}

\QualRule

% Header
\rule[-0.8ex]{0pt}{4.0ex}
{\large\bfseries Query-Relevant Passage (QRP)}
&
{\large\bfseries\color{nonsenseRed}
Generated NONSENSE String}
\\

\QualRule

% ------------------------------------------------------------
% SciFact
% ------------------------------------------------------------
\QualQuery{SciFact Query: Activator-inhibitor pairs are provided dorsally by Admpchordin.}

\QualQRP{
Regulation of ADMP and BMP2/4/7 at Opposite Embryonic Poles Generates a Self-Regulating Morphogenetic Field. Embryos have the ability to self-regu ...
}{BGE-base}{0.731}
&
\QualNonsense{
:::;$\times$ \};:: :;1: :::\{ |\&xa FN:p M::c :Uf: :|2v **P/ mdmf 1:\S{}, :*m- Mf1\& I|6, :*M: b\S{}:5 \%!\}] I*7] C:UB 2|2\S{} ]x:D \S{}\S{})6 7]H) a\S{}m: xR1\} 6|:: C ...
}{BGE-base}{0.870}
\\

\QualRule

% ------------------------------------------------------------
% SciDocs
% ------------------------------------------------------------
\QualQuery{SciDocs Query: Subjectivity and Sentiment Analysis of Modern Standard Arabic}

\QualQRP{
The Penn Arabic Treebank : Building a Large-Scale Annotated Arabic Corpus. From our three year experience of developing a large-scale corpus of an ...
}{BGE-base}{0.783}
&
\QualNonsense{
:::: ::0\{ -:*0 :\{:: *0:: \_A:] 5/0A C:C| j[c: ]*W1 :Z|\{ U0i\# \#\#\%- \%\%!\# [\#vO :uZL \&q3: r\#\_c ZS\{w Lf\#m nn\{Y M:\{T \%:lE :zs\{ \%-\#d m*W/ mT$\times$: D ...
}{BGE-base}{0.799}
\\

\QualRule

% ------------------------------------------------------------
% FIQA
% ------------------------------------------------------------
\QualQuery{FIQA Query: Something looks off about Mitsubishi financial data}

\QualQRP{
All but certainly, Mitsubishi is selling so cheaply because of the fuel scandal. It has been providing false fuel efficiency data for decades. As ...
}{BGE-base}{0.798}
&
\QualNonsense{
::*- [:[. FY\#2 (\&*\_ )E8: \#6z9 \_M\&t \textbackslash{}c;l ;]\}, \textbackslash{}C:? )xD) x9Zt )fy5 ?]\S{}, *:-* m.T: :\&\#2 -:-: *,*0 ((c? )2/\} \textbackslash{}J;W :\}fu w.:: --|\& -\#6m b;nN \} ...
}{BGE-base}{0.849}
\\

\QualRule

% ------------------------------------------------------------
% ArguAna
% ------------------------------------------------------------
\QualQuery{ArguAna Query: Syria clearly meets the standards for intervention The Assad regime has clearly lost its legitimacy and has pre ...}

\QualQRP{
global middle east house would arm syrian rebels. The if the rebels are armed and the regime gets close to losing surely it will simply increase t ...
}{BGE-base}{0.737}
&
\QualNonsense{
::1\# -\_i\# Q:\&. :\#\S{}\S{} d\_\_D =[/- \_\S{}*: =|,\_ |A\_* n1*= \S{}\}Cj :\S{}:2 \_A6: R4:v =4\&0 \_hb* ,2l\S{} s6+c )s,s 2.a7 \_-dv \_5=0 *\#:r 7\_,\_ +=RL 0=Rp */\%3 M ...
}{BGE-base}{0.814}
\\

\QualRule

% ------------------------------------------------------------
% NFCorpus
% ------------------------------------------------------------
\QualQuery{NFCorpus Query: Sexually Transmitted Fish Toxin}

\QualQRP{
Cluster of ciguatera fish poisoning--North Carolina, 2007.. Ciguatera fish poisoning (CFP) is a distinctive type of foodborne disease that results ...
}{BGE-base}{0.752}
&
\QualNonsense{
::*| -:Lf :c:s *k.: \}u;\& \&T!\} )C\#- :8|1 .A,5 7E47 \#\%8\& d978 5\&u= 6s1= 0:R- 9-6$\Delta$ K-k. LH\&P -Dv: bD|| NXfV :T:\& fo:L ;N67 /m!m b)]| ;(Y. 9 ...
}{BGE-base}{0.864}
\\

\QualRule

\end{tabular}
\endgroup

\vspace{-0.6mm}

% Table 9 — BGE
\caption{Representative BGE-base examples across five datasets. Each example shows a query, the retriever's highest-scoring judged query-relevant passage (QRP), and the optimized NONSENSE string, together with their dense similarity scores. Text is truncated for display.}

\label{tab:samples-bge}

\end{table*}

% ============================================================
% CONTRIEVER
% ============================================================

\begin{table*}[t]
\centering

\begingroup
\teaserfont
\footnotesize

\setlength{\tabcolsep}{7pt}
\setlength{\arrayrulewidth}{0.40pt}

\begin{tabular}{|
    >{\centering\arraybackslash}m{0.47\textwidth}|
    >{\centering\arraybackslash}m{0.47\textwidth}|
}

\QualRule

% Header
\rule[-0.8ex]{0pt}{4.0ex}
{\large\bfseries Query-Relevant Passage (QRP)}
&
{\large\bfseries\color{nonsenseRed}
Generated NONSENSE String}
\\

\QualRule

% ------------------------------------------------------------
% SciFact
% ------------------------------------------------------------
\QualQuery{SciFact Query: Activator-inhibitor pairs are provided dorsally by Admpchordin.}

\QualQRP{
Regulation of ADMP and BMP2/4/7 at Opposite Embryonic Poles Generates a Self-Regulating Morphogenetic Field. Embryos have the ability to self-regu ...
}{Contriever}{0.485}
&
\QualNonsense{
3S:a gl-A [Fa[ s?Q\} Qa=9 QSCD Q[Pk |pn\} .bd1 Ii=A nM/S \},y3 |Q;k dQa= 9|q| gCIL \}.y= 9[S1 |And (2)\_ 3\_V\_ 1\}\_5 |M\_8 |JbT -5-b q:8\_ 1\}y4 | ...
}{Contriever}{0.660}
\\

\QualRule

% ------------------------------------------------------------
% SciDocs
% ------------------------------------------------------------
\QualQuery{SciDocs Query: Subjectivity and Sentiment Analysis of Modern Standard Arabic}

\QualQRP{
Determining the Sentiment of Opinions. Identifying sentiments (the affective parts of opinions) is a challenging problem. We present a system that ...
}{Contriever}{0.482}
&
\QualNonsense{
vox/ J/@@ \},gQ t$\Delta$o\# \#t|S @|L\# A\}[a \#i/l a\#x| L\#f/ a\},[ i\#i\_ d/J* D/W1 j\#u| F\#D\# \#od8 f|EF |[M\# dkTu |\#F| lR\%\# D\}X\} \}OF( I!u\} $\times$1o\% \%|j\_ \} ...
}{Contriever}{0.664}
\\

\QualRule

% ------------------------------------------------------------
% FIQA
% ------------------------------------------------------------
\QualQuery{FIQA Query: Something looks off about Mitsubishi financial data}

\QualQRP{
All but certainly, Mitsubishi is selling so cheaply because of the fuel scandal. It has been providing false fuel efficiency data for decades. As ...
}{Contriever}{0.486}
&
\QualNonsense{
.*:: \_r9F J9U9 \_?S; S::; 4i$\times$$\mu$ 7M5$\times$ 7JW@ qN@\# GhDo ?s\#? R\&A\# ?B2B ?S@? l?M: :::: :::- /\_v\_ j3GW i::: :(() ((i! i?!O M::: 9:9\_ m?O: \_!-/ c ...
}{Contriever}{0.561}
\\

\QualRule

% ------------------------------------------------------------
% ArguAna
% ------------------------------------------------------------
\QualQuery{ArguAna Query: Syria clearly meets the standards for intervention The Assad regime has clearly lost its legitimacy and has pre ...}

\QualQRP{
global middle east house would arm syrian rebels. The if the rebels are armed and the regime gets close to losing surely it will simply increase t ...
}{Contriever}{0.449}
&
\QualNonsense{
OXmf \&T// EsC/ ./// /4]4 ]by/ vB0\_ !@=, rFIw R63\_ ?S2] ;X-8 f270 s2\&O ,,a0 Mg-? S3c, ]]]@ ;tZa ovWw \_/u\& iMF/ R\_;u s?s3 hB/\_ \_?r6 ?Q\}[ 4 ...
}{Contriever}{0.694}
\\

\QualRule

% ------------------------------------------------------------
% NFCorpus
% ------------------------------------------------------------
\QualQuery{NFCorpus Query: Sexually Transmitted Fish Toxin}

\QualQRP{
Ciguatera fish poisoning. A southern California epidemic.. Ciguatera fish poisoning results from the bioconcentration of a variety of toxins produ ...
}{Contriever}{0.523}
&
\QualNonsense{
.(Q: ;1(- 3\#3] \&]FV ;b5$\Delta$ Tsgv $\mu$v-9 ;\%34 899\} O|\}b FW\}7 A5\}7 bL/z o6\}F O(m9 ;K19 \}Eel [Sl] U17g |Z== \}8]\% 3|3\} 1\#e| J5\%| HFW| =]a8 =s|B T ...
}{Contriever}{0.640}
\\

\QualRule

\end{tabular}
\endgroup

\vspace{-0.6mm}

% Table 10 — Contriever
\caption{Representative Contriever examples across five datasets. Each example shows a query, the retriever's highest-scoring judged query-relevant passage (QRP), and the optimized NONSENSE string, together with their dense similarity scores. Text is truncated for display.}

\label{tab:samples-contriever}

\end{table*}

\end{document}